\documentclass[prl,reprint,superscriptaddress]{revtex4-1}

\usepackage{amsmath,amssymb,amsthm}
\usepackage{bbm,dsfont,mathrsfs,slashed,cancel} 
\usepackage{empheq}
\usepackage{graphicx} 
\usepackage{booktabs} %for tables
\usepackage{multirow} %for tables
\usepackage{float}
\usepackage{ytableau} %Young tableau
\ytableausetup{centertableaux}
\usepackage{tikz-cd} %commutative diagrams

\usepackage{upgreek}
\usepackage{bm}

\usepackage{verbatim}
\usepackage{hyperref}
\usepackage{color}

\newcommand {\cA}{{\cal A}}
\newcommand {\cB}{{\cal B}}

\newcommand {\cF}{{\cal F}}
\newcommand {\cG}{{\cal G}}
\newcommand {\cH}{{\cal H}}

\newcommand {\cJ}{{\cal J}}

\newcommand {\cL}{{\cal L}}
\newcommand {\cM}{{\cal M}}
\newcommand {\cN}{{\cal N}}

\def\a{\alpha}
\def\b{\beta}

\def\d{\delta}
\def\e{\epsilon}

\def\g{\gamma}

\def\l{\lambda}

\def\s{\sigma}

\def\F{\Phi}

\def\L{\Lambda}
\def\O{\Omega}

\def\S{\Sigma}
\def\U{\Upsilon}

\def\ri{{\rm i}}

\newcommand{\gd}{{\dot\g}}

\newcommand{\ad}{{\dot{\alpha}}}
\newcommand{\bd}{{\dot{\beta}}}

\newcommand{\sSU}{\mathsf{SU}}

\newcommand{\sU}{\mathsf{U}}

\newcommand{\ve}{\varepsilon}

\newcommand{\pa}{\partial}
\newcommand{\hf}{\frac12}

\newcommand{\be}{\begin{equation}}
	\newcommand{\ee}{\end{equation}}
\newcommand{\bea}{\begin{eqnarray}}
	\newcommand{\eea}{\end{eqnarray}}
\newcommand{\non}{\nonumber}
\newcommand{\ba}{\begin{array}}
	\newcommand{\ea}{\end{array}}

\def\double #1{#1{\hbox{\kern-2pt $#1$}}}

\newcommand{\ts}{{\tilde{\s}}}

\newcommand{\bsubeq}{\begin{subequations}}
	\newcommand{\esubeq}{\end{subequations}}

\newcommand{\rd}{\mathrm d}
\def \un{\underline}

\newcommand{\de}{{\nabla}}
\newcommand{\deb}{{\bar{\nabla}}}

\newcommand{\loco}{\vert}
\newcommand{\doubar}{{{\loco}\!{\loco}}}

\newcommand{\body}{\big|}
\newcommand{\dbody}{\big|\!\big|}
\newcommand{\tproj}{\big|\!\big|\!\big|}

\newcommand{\ua}{\underline{a}}
\newcommand{\ub}{\underline{b}}
\newcommand{\uc}{\underline{c}}

\def\a{\alpha}
\def\b{\beta}
\def\g{\gamma}

\def\d{\delta}

\def\e{\epsilon}
\def\ve{\varepsilon}

\def\F{\Phi}

\def\l{\lambda}
\def\L{\Lambda}

\def\s{\sigma}
\def\S{\Sigma}

\def\O{\Omega}

\newcommand{\doihref}[2]{\href{https://doi.org/#1}{#2}}

\newcommand{\arxivlink}[1]{\href{https://arxiv.org/abs/#1}{{\ttfamily arXiv:#1 [hep-th]}}}

\begin{document}

	\title{Superform Approach to Equivariant Localization in Supergravity}

	\author{Michele Galli}
	\email{m.galli@uq.edu.au}
	\affiliation{School of Mathematics and Physics, University of Queensland, 
		St Lucia, Brisbane, Queensland 4072, Australia}
	
	\author{Christian Kennedy}
	\email{christian.kennedy@uq.edu.au}
	\affiliation{School of Mathematics and Physics, University of Queensland, 
		St Lucia, Brisbane, Queensland 4072, Australia}
	
	\author{Parth Raina}
	\email{p.raina@uq.edu.au}
	\affiliation{School of Mathematics and Physics, University of Queensland, 
		St Lucia, Brisbane, Queensland 4072, Australia}

	\author{Gabriele Tartaglino-Mazzucchelli}
	\email{g.tartaglino-mazzucchelli@uq.edu.au}
	\affiliation{School of Mathematics and Physics, University of Queensland, 
		St Lucia, Brisbane, Queensland 4072, Australia}

	\date{\today}

	\begin{abstract}

We identify a superspace mechanism behind equivariant localization in supergravity. We show that closed superforms generate, on supersymmetric backgrounds, equivariantly closed polyforms. After presenting the general mechanism, we construct such polyforms for vector and linear multiplets, and for chiral and BF action principles, in off-shell \(4d\) \(\cN=2\) conformal supergravity, reproducing and extending recent results. Our construction provides a geometric first step toward equivariant localization of BPS observables in supergravity, including higher-derivative theories, and holography.

	\end{abstract}
	
 	\maketitle
	\allowdisplaybreaks

\section{Introduction} 
%\textit{Introduction.}---

Supersymmetric localization represents one of the most remarkable developments in supersymmetric quantum field theory. By reducing certain infinite-dimensional path integrals to finite-dimensional integrals or sums, it has enabled exact non-perturbative results, with applications to string theory, precision tests of holographic (AdS/CFT) dualities, and even connections to geometry and topology. The literature is vast; see the comprehensive review \cite{Pestun:2016zxk} and references therein.

Supersymmetric localization is closely linked to equivariant localization. A preserved supercharge squares to a symmetry generated by a Killing vector \(\xi\). Equivariant localization is governed by the differential \(\rd_\xi=\rd-\iota_\xi\) and its fixed-point formulae \cite{Atiyah:1984px,Berline:1982,Pestun:2016zxk}. Recently, the systematic discovery of \(\rd_\xi\)-closed polyforms associated with BPS observables in supergravity, including (on-shell) actions, has led to new exact results in holography \cite{BenettiGenolini:2023kxp,Martelli:2023oqk}; see also \cite{Dabholkar:2010uh,Hristov:2018lod,Hosseini:2019iad,Hristov:2021qsw,Hristov:2022lcw,Hristov:2022plc,Colombo:2023fhu,Hristov:2024cgj,BenettiGenolini:2024xeo,Cassani:2024kjn,BenettiGenolini:2024lbj,Colombo:2025ihp,Colombo:2025yqy,Gaar:2026nqq,BenettiGenolini:2026hmz,BenettiGenolini:2026qdm,BenettiGenolini:2024kyy,BenettiGenolini:2025icr,Arav:2025jee,Park:2025fon,Arav:2026unc,Cassani:2026teb,Couzens:2026xmi,BenettiGenolini:2026cyc,BenettiGenolini:2026cdw}.
However, the origin of these equivariantly closed polyforms in supergravity has not been explained so far.

A natural setting for this question is superspace. In supergravity, local supersymmetric invariants can be constructed from closed supersymmetric differential forms, or superforms; see \cite{Castellani:1991eu,Gates:1997kr,Gates:1997ag,Gates:1998hy} for seminal works and \cite{Howe:2003cy,Berkovits:2006ik,Berkovits:2008qw,Gates:2009uv,Gates:2009xt,Butter:2012ze,Kuzenko:2012ew,Butter:2013rba,Kuzenko:2013rna,Kuzenko:2014jra,Arias:2014ona,Castellani:2014goa,Gates:2014cqa,Linch:2014iza,Castellani:2015paa,Butter:2016qkx,Castellani:2016ibp,Becker:2017njd,Butter:2018wss,Butter:2019edc,Cremonini:2021vyy,Grassi:2023bxm,Grassi:2023yfe,Soueres:2016qre,Howe:2011tm} for further developments, including applications to higher-derivative supergravity. Since both closed superforms and equivariantly closed polyforms encode supersymmetric Lagrangians, it is natural to ask whether the latter arise directly from the former.

We show that this expectation is naturally realized. The component-field projection of a closed superform contains the gravitino expansion of supergravity fields and invariants. On a supersymmetric background, appropriately replacing the gravitini by the Killing spinors of the preserved supersymmetry produces an equivariantly closed polyform. In superspace, this replacement is implemented by contracting the superform with the odd vector superfield generating the preserved supersymmetry. This prescription is the main result of our paper.

We apply our proposal to off-shell \(4d\) \(\cN=2\) conformal supergravity, a central setting for effective supergravity and precision holography. Recent proposals and applications of equivariant localization in $4d$ $\cN=2$ higher-derivative supergravity appeared in \cite{Hristov:2022lcw,Hristov:2024cgj,BenettiGenolini:2026qdm}. We construct the polyforms for vector and linear multiplets, chiral Lagrangians, and BF couplings, thereby recovering and extending results in \cite{BenettiGenolini:2026qdm}. Our work gives a geometric explanation for the existence of equivariantly closed polyforms in supergravity and suggests a systematic route to new structures and applications in this context.

%%%%%%%%%%%%%%%%%%%%%%%%%%%%%%
\section{Superspace and Superforms: a primer}	
%\textit{Superspace and Superforms: a primer}---

We consider a curved superspace $\cM \!\equiv\! \cM^{d|\d}$ with $d$ bosonic and $\d$ fermionic dimensions. The local coordinates parametrizing  $\cM$ are $z^M\!=\!(x^m,\theta^{\hat{\mu}})$ with $x^m$ commuting and $\theta^{\hat{\mu}}$ Grassmann variables with appropriate spinorial and R-symmetry indices. See \cite{SUPERSPACE,Wess:1992cp,Buchbinder-Kuzenko,DeWitt:2012mdz,Kuzenko:2022skv,Kuzenko:2022ajd} for reviews of curved superspace and supergravity.
The coordinates carry Grassmann parities $|x^m|\!=\!|m|\!=\!0$ (even) and $|\theta^{\hat{\mu}}|\!=\!|\hat{\mu}|\!=\!1$ (odd).
We employ a local tangent space frame with supervielbein $E^A\!:=\! \rd z^M E_M{}^A(z)$, $E_A \!:=\! E_A{}^M(z) \partial_M$, where $E_A{}^M E_M{}^B \!=\! \d_A^B$ and $E_M{}^A E_A{}^N \!=\! \d_M^N$. The parities of these superfields are chosen so that $E^A,E_A$ are odd when $A$ is a spinorial index and even otherwise, i.e.~$|E_M{}^A| \!=\! |M| \!+\! |A|$. As in ordinary geometry, taking tensor products of $\rd z^M,\partial_M$ or of $E^A,E_A$ gives a basis to express components of general tensor superfields.

The superspace also has a connection super 1-form, $\Phi = E^A \Phi_A{}^{\ub} X_{\ub}$. The generators $X_{\ub}$ of the structure group include those of the Lorentz algebra and potentially others like R-symmetry, collectively labeled by an index $\un{a}$. This leads to a covariant derivative
\bea
&\nabla_A =E_A -\F_A
=E_A{}^M(\pa_M-\Phi_M{}^{\ub}X_{\ub})~.
\eea
The curved superspace $\cM$ is, up to some deformation, a Cartan geometry modeled on flat superspace.
This means that, if \(X_{\ua} U_{M_1\dots M_m}{}^{N_1\dots N_n}=0\), then \(U_{A_1\dots A_m}{}^{B_1\dots B_n}\) inherits the action of the \(X_{\ua}\) generators, such as the Lorentz generators, from the supervielbein \(E^A\) and its dual \(E_A\). This is the superspace analog of Riemannian geometry in a local frame.
The covariant derivatives satisfy the graded commutation relations
\bea
[\nabla_A , \nabla_B \} = -T_{AB}{}^C \nabla_C - R_{AB} ~, ~~R_{AB}:=R_{AB}{}^{\uc} X_{\uc}\,,~
\label{curved-algebra-1}
\eea
with $T_{AB}{}^C$ the torsion, and $R_{AB}$ the curvature of $\cM$. 

Grassmann parity means that \eqref{curved-algebra-1} is an anticommutator if both $A$ and $B$ indices are spinorial, and a commutator otherwise. Infinitesimal superdiffeomorphisms, or general coordinate transformations in superspace, are generated by vector superfields ${\mathscr X}:= {\mathscr X}^M \partial_M = {\mathscr X}^A E_A$ through Lie derivatives. 
A generic tensor superfield $U$ can be intrinsically odd $|U|=1$ or be the sum of even and odd parts. Taking a ${\mathscr X}$ with definite parity as an example, its components will have parity $|{\mathscr X}^A| = |{\mathscr X}| + |A|$.

Differential forms in superspace (superforms) behave, up to signs, much as in ordinary geometry.
In particular, Cartan's formula
$\cL_{\mathscr X}=\rd\iota_{\mathscr X}+\iota_{\mathscr X}\rd$
continues to hold for a vector field \({\mathscr X}\) of arbitrary parity. Infinitesimal superdiffeomorphisms act as $\d_{\mathscr X}\Omega=\cL_{\mathscr X}\Omega$ on a superform $\Omega$.
Our conventions for \(\rd\), \(\iota_{\mathscr X}\), and wedge products are collected in the Supplemental Material.

Importantly, $\rd^2=0$ remains true in superspace allowing for an analog of de Rham cohomology \cite{Gates:1980ay}. However, the interior product, given on super 1-forms by $\iota_{{\mathscr X}} E^A \!=\!{\mathscr X}^A$, has properties depending on parity: $\iota_{\mathscr X} \iota_{{\mathscr Y}} = -(-1)^{|{\mathscr X}||{\mathscr Y}|} \iota_{{\mathscr Y}}\iota_{\mathscr X}$. 
For an odd vector field $\e \!=\! \e^a E_a \!+\! \e^{\hat{\a}}E_{\hat{\a}}$, with $\e^a$ anticommuting and  $\e^{\hat{\a}}$ commuting, this implies that $\iota_\e^2 \neq 0$.
For example, on an ordinary manifold, a Killing spinor is taken to be commuting. We exploit this in the next section. 

If a super $p$-form $\Omega$ is invariant under the structure group, $X_{\ua}\Omega=0$, then the exterior derivative coincides with the exterior covariant derivative $\rd \Omega= \nabla \Omega$. Noting that the torsion is given by $T^A = \nabla E^A$, this gives 
\bea
(\rd \Omega)_{BA_1\cdots A_p}
&=&
(p+1)\nabla_{[B}\Omega_{A_1\cdots A_p)}
\non\\
&&
+\frac{p(p+1)}{2}T_{[BA_1}{}^{C}\Omega_{|C|A_2\cdots A_p)}
\,,
\label{closure-superform}
\eea
where $[\dots)$ is the graded antisymmetrization. We denote by \(J\) a closed superform, \(\rd J=0\). The dimension-zero torsion of superspace, inherited from the flat supersymmetry algebra and proportional to gamma matrices, organizes \eqref{closure-superform} into a cohomological hierarchy where components of \(J\) with fewer spinor indices are determined by derivatives of those with more spinor indices; see, e.g., \cite{Gates:1980ay,Gates:1997kr,DeWitt:2012mdz,Kuzenko:2022ajd,Kuzenko:2022skv,Gates:1997ag,Buchbinder-Kuzenko,SUPERSPACE,Wess:1992cp,Gates:1998hy,Cederwall:2001dx,Howe:2003cy,Berkovits:2006ik,Berkovits:2008qw,Gates:2009uv,Gates:2009xt,Greitz:2011da,Butter:2012ze,Kuzenko:2012ew,Butter:2013rba,Kuzenko:2014jra,Arias:2014ona,Castellani:2014goa,Gates:2014cqa,Linch:2014iza,Butter:2016qkx,Castellani:2016ibp,Becker:2017njd,Butter:2018wss,Butter:2019edc,Kennedy:2025nzm,Bandos:1997ui,Bergshoeff:1996tu,Howe:2011tm,Kuzenko:2013rna,Soueres:2016qre}.

Given a superfield $U(z)$, the projection to the bosonic body $\cM^d$ of $\cM^{d|\d}$ is denoted by the bar projection $U(x):=U(z)\loco=U(z)\loco_{\theta=0}$. Superforms are projected onto $\cM^d$ using the double-bar projection, which sets $\theta^{\hat{\a}}=0$ and $\rd \theta^{\hat{\a}}=0$. Then, $e^a=\rd x^m e_m{}^a=E^a\doubar$ is the standard vielbein while $\psi^{\hat{\a}}=\rd x^m \psi_m{}^{\hat{\a}}=2E^{\hat{\a}}\doubar$ is the gravitino field of a supergravity associated to the geometry of \eqref{curved-algebra-1}. 
Double-bar projecting a superform leads to a standard $p$-form with an expansion in gravitini:
\bsubeq
\label{gravitini-expansion}
\bea
\varOmega
&:=&\Omega\dbody
=
\sum_{q=0}^{p}
\frac{(-1)^{q(p-q)}}{2^{q}(p-q)! q!}\,
\varOmega^{(p-q)} 
\, ,~~~~
\label{gravitini-expansion-1}
\\
\varOmega^{(r)}&:=&e^{a_{r}}\cdots e^{a_1}
\psi^{\hat\alpha_{p-r}}\cdots\psi^{\hat\alpha_1}
\,
\varOmega_{a_1\cdots a_{r}\hat\alpha_1\cdots\hat\alpha_{p-r}}
\,.~~~~
\label{gravitini-expansion-2}
\eea
\esubeq
Here we omitted wedge products between forms and $\varOmega_{a_1\cdots a_{r}\hat\alpha_1\cdots\hat\alpha_{p-r}}:=\Omega_{a_1\cdots a_{r}\hat\alpha_1\cdots\hat\alpha_{p-r}}\loco$.

For a closed ${\rm rank}(J)=d$ superform, also invariant under the structure group, $X_{\ua}J=0$ \cite{note1}, the integral 
\bea
\int_{\cM^d}\mathcal{J}
\,,~~~~~~\mathcal{J}=J\doubar 
\,,
\label{ectoplasm}
\eea
is an invariant under all supergravity gauge transformations \cite{Castellani:1991eu,Gates:1997kr,Gates:1997ag,Gates:1998hy}. Superdiffeomorphism invariance derives from the fact that, with $\rd J=0$, 
$\d_{\mathscr X} J=\cL_{\mathscr X} J=\rd\iota_{\mathscr X} J$, which is a total derivative. Eq.~\eqref{gravitini-expansion} shows how the expansion in gravitini of supergravity actions naturally emerges from \eqref{ectoplasm}. Next, we show how to obtain an equivariantly closed polyform from a closed superform by appropriately replacing the gravitini fields in \eqref{gravitini-expansion} with Killing spinors.

%%%%%%%%%%%%%%%%%%%%%%%%%%%%%%%%%

\section{From superforms to equivariantly closed polyforms: generalities}	
%\textit{From superforms to equivariantly closed polyforms: generalities}---

There is one key formula in our work \cite{note2}.
On any supermanifold \(\cM\), for an odd vector field \(\e\), one has
\begin{equation}\label{eq:expdexp}
e^{\iota_\e}\rd e^{\iota_\e} = \rd  + \frac{1}{2} \iota_{\{\e,\e\}}+ \cL_\e
\,.
\end{equation}
The proof is simple and uses the natural properties of $\rd$, $\iota_\e$, and $\cL_\e$ on ordinary manifolds uplifted to supermanifolds. These properties are as follows. For any vector fields ${\mathscr X}$, ${\mathscr Y}$ of parities $|{\mathscr X}|$ and $|{\mathscr Y}|$, the following equations hold when acting on differential forms
\bea
&\cL_{\mathscr X} =  \{\rd,\iota_{\mathscr X}\}
\,,~~~
\iota_{[{\mathscr X},{\mathscr Y}\}} = [\cL_{\mathscr X}, \iota_{\mathscr Y}\} = [\iota_{\mathscr X}, \cL_{\mathscr Y}\}
\,,~~~~
\eea
with $[\,,\}$  the graded commutator.
Then, taking $\cA,\cB$ to be indeterminate, one has the series of anticommutators 
\bea
e^\cA\cB e^\cA = 
e^{\{\cA,~\}}\cB&=&
\cB 
+ \{\cB,\cA\} 
+ \frac{1}{2} \{\{\cB,\cA\},\cA\} 
\non\\
&&
+\frac{1}{6} \{\{\{\cB,\cA\},\cA\},\cA\}
+ \dots
\label{eABeA}
\eea
Replacing $\cA$ by $\iota_\e$ and $\cB$ by $\rd$, when $|\e|=1$ the equations
$\{\{\{\rd,\iota_\e\},\iota_\e\},\iota_\e\} = \{\iota_{\{\e,\e\}}, \iota_\e\} = 0$  hold. Thus, the second line of \eqref{eABeA} is zero, and \eqref{eq:expdexp} follows immediately. 

Using  \eqref{eq:expdexp}, and taking a generic super $p$-form $\O$, one obtains another fundamental equation: 
\begin{equation}\label{eq:dexpO}
\left(
\rd 
-\iota_{\frac{1}{2}\{\e,\e\}}
\right)
e^{\iota_\e}\O 
=e^{-\iota_\e}\left( \rd  + \cL_\e\right)\O
\,.
\end{equation}
We stress that \eqref{eq:dexpO} holds for any odd vector field $\e$ and superform $\O$. Applying \eqref{eq:dexpO} to a closed ($\rd J=0$), and $\e$-invariant ($\cL_\e J=0$) super $p$-form $J$,  we obtain
\bea
&\left(
\rd 
-\iota_{\Xi}
\right)
J(\e)
=0
\,,~~~~
\Xi:=\hf\{\e,\e\}\,,~
J(\e):=e^{\iota_\e}J
\,.~~~~
\label{eq:dexpJ0b}
\eea
Since $\iota_\e^nJ$ is a super $(p-n)$-form,  $J(\e)$ is a supersymmetric polyform and, remarkably, \eqref{eq:dexpJ0b} looks like a supersymmetric equivariant closure condition. 

It is very natural to argue that by double-bar projecting $J(\e)$, and choosing $\e$ to be an odd vector field such that $\e^{\hat{\a}}\loco$ is a Killing spinor, the following should hold
\begin{align}\label{eq:SuperspaceEquivariantForm}
\cJ(\e):=J(\e)\tproj
\,,~~~
\big(\rd- \iota_\xi \big) \cJ(\e)
= 0
\,,
\end{align}
with $\xi^a\propto\bar{\e} \Gamma^a \e$ a Killing vector constructed from two commuting Killing spinors.
Eq.~\eqref{eq:SuperspaceEquivariantForm} is the equivariant closure condition on the polyform $\cJ(\e)$. 
Note that, with the triple-bar projection in \eqref{eq:SuperspaceEquivariantForm}, we indicate taking the double-bar projection of $J$ together with restricting to field configurations invariant under supersymmetry, with all fermionic fields set to zero, e.g., $\psi_m{}^{\hat{\a}}=0$. 

For general superspace discussions of (conformal) super-isometries of supergravity backgrounds, see \cite{Buchbinder-Kuzenko, Kuzenko:2015lca}.
Now, it is not hard to argue that \eqref{eq:SuperspaceEquivariantForm} follows from \eqref{eq:dexpJ0b} under fairly general assumptions on the underlying supergeometry and on its superisometries. 
First of all, note that with the odd vector field $\e$, we aim to describe only fermionic transformations within the set of superisometries. In this case, $\e^{\hat{\alpha}}\loco$ is identified with the Killing spinor and $\e^a\propto(\bar\theta\Gamma^a\e)+\cdots$ leading to $\e^a\loco=0$. By using the superspace version of the Killing spinor equation
\be
\d_\e E^A = \nabla \e^A + \iota_\e T^A + \Lambda E^A = 0
\,,
\label{Superspace-Killing-equation}
\ee
with  $\Lambda=\Lambda^{\underline{a}}X_{\underline{a}}$
an appropriate structure group gauge transformation, and using $\iota_\e(\Lambda E^a)\loco=0$, one then obtains
\bea
\Xi\tproj
=
-\frac{1}{2}\iota^2_\e T^AE_A\loco
=\e^{\hat{\g}}\e^{\hat{\b}}T_{\hat{\b}\hat{\g}}{}^aE_a\loco
\propto
(\ri\bar{\e}\Gamma^a\e)\,e_a{}^m\pa_m
\,.~~
\eea
Here, the gamma-matrix emerges from the standard mass dimension zero torsion of superalgebras, $T_{\hat{\a}\hat{\b}}{}^c\propto(\Gamma^c)_{\hat{\a}\hat{\b}}$, and, for simplicity, we have assumed that the only dimension-zero torsion is the flat one \cite{note4}.

Let us stress that we expect these arguments to be general, applying to both on-shell and off-shell supergravities. We also expect \eqref{eq:SuperspaceEquivariantForm} to admit a direct component-field derivation. Supergravity observables, such as Lagrangian forms, possess gravitino expansions as in \eqref{gravitini-expansion}, independently of a superspace construction. According to \eqref{eq:SuperspaceEquivariantForm}, on a supersymmetric bosonic background, equivariantly closed polyforms are obtained from \eqref{gravitini-expansion} by replacing (up to appropriate factors of two and signs) gravitino one-forms with commuting Killing spinors:
\bsubeq
\label{Killing-spinor-expansion}
\bea
\cJ(\e)
\!&=&\!
 \sum_{s=0}^{\lfloor p/2\rfloor}(-1)^{s}\,\cJ^{(p-2s)}(\e)
 \,,~~~~
\label{Killing-spinor-expansion-1}
\\
\cJ_{a_1\cdots a_{r}}^{(r)}(\e)
\!&=&\!
 \frac{1}{(p-r)!}\,\cJ_{a_1\cdots a_{r}\hat\alpha_1\cdots\hat\alpha_{p-r}}\e^{\hat\alpha_1}\cdots\e^{\hat\alpha_{p-r}}
\,.~~~~~~~
\label{Killing-spinor-expansion-2}
\eea
\esubeq
Here 
$\cJ^{(r)}(\e)\!:=\!\frac{1}{r!}e^{a_{r}}\wedge\cdots\wedge e^{a_1} \,\cJ_{a_1\cdots a_{r}}^{(r)}(\e)$, 
we identified $\e^{\hat{\a}}$ with $\e^{\hat{\a}}\loco$,
$\cJ_{a_1\cdots a_{r}\hat\alpha_1\cdots\hat\alpha_{p-r}}=J_{a_1\cdots a_{r}\hat\alpha_1\cdots\hat\alpha_{p-r}}\loco$, and in \eqref{Killing-spinor-expansion-1} we used that, on a bosonic background with no background fermionic fields,
$J_{a_1\cdots a_r\hat\alpha_1\cdots\hat\alpha_{p-r}}\body\!=\!0$ whenever $(p-r)$ is odd.
Eq.~\eqref{Killing-spinor-expansion} provides an ansatz for equivariantly closed polyforms directly in components from supersymmetric completions, justified by \eqref{eq:dexpO}--\eqref{eq:SuperspaceEquivariantForm}.

One further comment is worth making. In practice, \(\cL_\e J=0\) amounts to imposing the supersymmetry conditions on $J$'s component fields. When \(J\) contains bare connection forms, as in Chern--Simons-type couplings, this requirement may also impose a gauge condition; the BF example below illustrates this point.

%%%%%%%%%%%%
\section{The $4d$ $\cN=2$ case}
%\textit{The $4d$ $\cN=2$ case}---

We focus on $4d$ $\cN=2$ conformal supergravity and its supersymmetric backgrounds. We refer the reader to \cite{deWit:1979dzm,deWit:1980lyi,deWit:1980gt,deWit:1982na,deWit:1984wbb,deWit:1984rvr,Freedman:2012zz,Lauria:2020rhc} for standard references on the superconformal tensor calculus for $4d$ $\cN=2$, and \cite{Butter:2009cp,Butter:2011sr,Butter:2012xg,Kuzenko:2022ajd,Gold:2024gsj} for the so-called conformal superspace where the $\cN=2$ superconformal algebra is gauged off-shell in superspace. See \cite{BenettiGenolini:2026qdm,Klare:2012gn,Cassani:2012ri,Klare:2013dka,Butter:2015tra,Kuzenko:2015lca} for the study of supersymmetric backgrounds in $4d$ conformal supergravity.

In terms of component fields, the Weyl multiplet of $\cN=2$ conformal supergravity comprises the independent gauge fields $\{e_{m}{}^{a}, {\psi_{m}}_{i}^{\a}, \bar\psi_{m \ad}^{\phantom{b}\phantom{b} i}, b_{m}, A_{m}, \phi_{m}{}^{i j} \}$, together with covariant matter fields $\{W_{a b}, \S_{\a}^{i}, \bar\S_{i}^{\ad}, D\}$. 
Here, $b_{m}$ is the dilatation gauge field, $A_{m}$ and $\phi_{m}{}^{i j}$ are $\sU(1)_{R}$ and $\sSU(2)_{R}$ gauge fields, $W_{a b}$ is a real anti-symmetric tensor with (anti-)self dual components $W_{a b}^{\pm}$, and $D$ is a real scalar.
We work in Minkowski signature following \cite{Butter:2012xg,Gold:2024gsj}. However, our analysis can be extended to Euclidean signature, which is the playground for applying equivariant localization theorems, where, as in \cite{deWit:2017cle,BenettiGenolini:2026qdm}, spinors would be left/right chirality ${\sSU}(2)$-Majorana-Weyl, and the Killing-spinor bilinears are real.

The underlying conformal superspace uses covariant derivatives $\de_A=E_A-\omega_A{}^{\uc} X_{\uc}$ 
where $X_{\uc}$ includes all generators of the $\cN=2$ superconformal algebra except for the momenta and $Q$-supersymmetry generators, that are effectively described by the $\de_a$ and $(\de_\a^i,\bar{\de}^\ad_i)$ derivatives.

We are interested in $\cN=2$ conformal supergravity, coupled to vector and linear matter multiplets, see \cite{Butter:2012xg,Gold:2024gsj}.
The Abelian vector multiplet is described by a real gauge super 1-form $V$ with closed superfield strength $F = \rd V$, $\rd F=0$. 
In components the multiplet comprises the following fields $\{\phi,\,\bar{\phi},\, \lambda^i_\a,\, \bar\l_{i}^{\ad}, \, X^{ij},\, f_{ab}\}$, with $f_{ab}=2e_a{}^me_b{}^n\pa_{[m} v_{n]}$ and $X^{ij}=X^{ji}$ satisfying a reality condition $(X^{ij})^*=\ve_{ik}\ve_{jl}X^{kl}=X_{ij}$. 
The linear multiplet is described in superspace by a real gauge super 2-form $B$ with  closed super field strength $H=\rd B$, $\rd H=0$.
It has the following components $\{G^{ij},\, \chi^i_\a,\,\bar{\chi}_i^\ad,\, F,\,\bar{F},\, h_{abc}\}$ with $h_{abc}=3e_a{}^me_b{}^ne_c{}^p\pa_{[m}b_{np]}$ and $G^{ij}=G^{ji}=(G_{ij})^*$. We will also consider a generic chiral multiplet describing a supersymmetric Lagrangian, and a BF Lagrangian coupling, for which some details will be presented later.

We enforce a supersymmetric background on $\cM^d$ by setting the combined $Q$- and $S$-supersymmetry transformations of all multiplets to zero. 
In superspace, the odd Killing vector field is $\e =\e^aE_a+ \e^\a_i E_\a^i + \bar{\e}^\ad_i \bar{E}_\ad^i$, and the projected covariant fermionic transformation is \cite{Freedman:2012zz,Lauria:2020rhc,Butter:2011sr,Kuzenko:2022ajd}
\begin{align}
\d_\e^{\rm covariant}=\e^\a_i Q_\a^i + \bar{\e}^\ad_i \bar{Q}_\ad^i
+\eta_\a^i S^\a_i+\bar{\eta}^\ad_i \bar{S}_\ad^i
\,,
\end{align}
with $(S^\a_i,\bar{S}_\ad^i)$ generating $S$-supersymmetry, $Q_\a^i=\de_\a^i\loco$, $\bar{Q}^\ad_i=\bar{\de}^\ad_i\loco$, and we used $\e^a\loco=0$.
Setting the gravitini variation to zero, $\d\psi_{i}=\d\bar\psi^{i}=0$, provides the Killing spinor equation (we use spinor contractions as in \cite{Buchbinder-Kuzenko})
\be
    2\nabla_{a}\e_{i}^{\a} - \frac{\ri}{2} (\bar \e_{i} \tilde\s_{a} \s^{c d})^{\a} W_{c d} + 2\ri (\bar\eta_{i}\tilde \s_{a})^{\a} = 0 \, ,
    \label{eq:gravitino-variation}
\end{equation}
together with its conjugate. 
We abuse notation denoting $\e^\a_i\loco$ as $\e^\a_i$ and similarly for $\bar{\e}^i_\ad$, the derivative $\de_a=\de_a\loco$ and the commuting $S$-supersymmetry parameters $\eta^{i}_\a$, $\bar\eta^\ad_{i}$.
Imposing the zero-gaugino supersymmetric background condition, $\d\l_\a^i=\d\bar{\l}^\ad_i=0$, implies
\bea
0&=& 
2 (\s^{a b} \e^{i})_{\a} f_{a b} 
- \frac{1}{4} (\s^{a b} \e^{i})_{\a} W_{a b} \bar\phi  
- \frac{1}{2} \e_{\a j} X^{i j} 
\non\\
&&
+ 2\ri (\s^{a} \bar \e^{i})_{\a} \nabla_{a} \phi 
+ 4\eta_{\a}^{i} \phi
\, , 
    \label{eq:gaugino-variation}
\eea
together with its conjugate.
For the linear multiplet, supersymmetry invariance $\d\chi_{\a}^{i} = \d\bar\chi_{i}^{\ad} = 0$ leads to
\be
     \e _{\a i}F
     +4 \ri (\s ^a \bar\e _i)_\a h_a 
     - \ri (\s ^a \bar \e ^j)_\a\nabla _a G _{ij}
     - 4 \eta _\a^j G_{ji} = 0 \, ,
\label{eq:chi-variation}
\ee
together with its conjugate. Here $h_{a} = \frac{1}{6} \ve_{abcd}h^{bcd}$ is the Hodge dual of the closed 3-form $h_{abc}$.

We now construct the polyforms for the vector and linear multiplets, $\cF(\e):=e^{\iota_\e} F\tproj$,
$\cH(\e):=e^{\iota_\e} H\tproj$. The reader should be careful to consider the interior product of both the $\e^\a_i$ and $\bar{\e}_\ad^i$ Killing spinors in the odd vector field $\e$. By using the expressions for $F$ and $H$ from \cite{Butter:2012xg,Gold:2024gsj}, given also in the supplemental material, we obtain
\be
\cF(\e)
=
f
+ S\bar \phi 
- \bar{S}  \phi 
\,,~~~
\cH(\e)
=
h
-\frac{1}{2} K^{ij}G_{i j}
\,,~~~
\ee 
where $f=\frac{1}{2} e^b \wedge e^a f_{a b} 
$, $h=\frac{1}{6} e^{c} \wedge e^{b} \wedge e^{a} h_{abc} $, and the bilinear scalars and 1-form are 
\be
S:=(\e_i\e^i)
\,,~~~
\bar{S}:=(\bar\e^i\bar\e_i)
\,,~~~
K^{ij}
:=e^a(\e^{(i}\s_{a}\bar\e^{j)}) 
\,.
\ee
Using \eqref{eq:gravitino-variation}, \eqref{eq:gaugino-variation}, \eqref{eq:chi-variation}, and their conjugates, one can show that $\cF(\e)$ and $\cH(\e)$ are equivariantly closed polyforms with respect to a Killing vector $\xi$
which, for $4d$ $\cN=2$, is
\begin{align}
\frac{1}{2}\{\e,\e\}\tproj=\xi=\xi^a e_a{}^m\pa_m
\,,~~~
\xi^a = -2\ri (\e_i \sigma^a\bar{\e}^i)
\,.
\end{align}

A vector multiplet on a generic background can satisfy an on-shell condition $X^{ij}=0$ \cite{note3}.  In superspace, this means that $F={\mathscr P}+\bar{{\mathscr P}}$, for complex closed super 2-forms
${\mathscr P}$ and $\bar{\mathscr P}$; $\rd{\mathscr P}=\rd\bar{\mathscr P}=0$ \cite{Gates:2009xt,Butter:2017pbp}. The corresponding polyforms are 
\be
 \cG(\e)= \e^{\iota_\e}{\mathscr P}\tproj
 = g
 +S\bar \phi \, ,~~~~
\bar{\cG}(\e)
= \e^{\iota_\e}\bar{{\mathscr P}}\tproj=
\bar{g}
- {\bar S}\phi 
\, .
\ee
Here $g\!=\!\frac{1}{2} e^{b} \wedge e^{a} g_{a b}$, 
$g_{a b} \!=\!  f_{ab}^{-} - \hf W_{a b}^{+} \bar \phi +\hf W_{ab}^{-} \phi$,
such that $\rd g=0$, together with its conjugate.
Using \eqref{eq:gravitino-variation}, \eqref{eq:gaugino-variation} and their conjugates, with $f_{ab}$ appropriately replaced by $g_{ab}$ and $\bar{g}_{ab}$, one can prove the independent equivariant closure equations 
$(\rd-\iota_\xi) \cG(\epsilon)=0$ and $(\rd-\iota_\xi) \bar{\cG}(\epsilon)=0$.

An important result of \cite{BenettiGenolini:2026qdm} was to construct an equivariantly closed polyform from the chiral ($F$-term) Lagrangian, which can be used to describe a wealth of models, including higher-derivative supergravities. We take the closed super 4-form $J^{{\rm chiral}}$ associated with a chiral action \cite{Gates:2009uv,Kuzenko:2022ajd} (see also the supplemental material) and construct the polyform 
$\cJ^{{\rm chiral}}(\e)=e^{\iota_\e}J^{{\rm chiral}}\tproj$:
\bsubeq\label{chiral-polyform}
\bea
\cJ^{{\rm chiral}}(\e)
&=&
\left(\frac{1}{2} C + \frac{1}{2} W^{-}{}^{ab}W^{-}_{ab}\, \Phi\right)
{\rm vol}_4
+2 \ri \bar{\U}^{ij}B_{i j}
\non \\
&&
+2\ri \left(
 \bold{F}^+
+  2 W^-\Phi\right)
\bar{S}
+ 4\ri\Phi\,\bar{S}^2 
\,,
\\
\bar{\Upsilon}^{ij}
&:=&
-\frac{1}{2} e^b\wedge e^a(\bar\e^{i} \tilde \s_{a b}\bar\e^{j})
\,.
\eea\esubeq
Here ${\rm vol}_4=\frac{1}{24} \ve_{abcd} e^{d} \wedge e^{c} \wedge e^{b} \wedge e^{a} $, $\bold{F}^+=\frac{1}{2} e^b \wedge e^a \bold{F}^+_{a b}$, $W^-=\frac{1}{2} e^b \wedge e^a W^-_{a b}$,
and $\Phi=\cL_c$ is a dimension two superconformal primary, while $B_{i j}=B_{ji}$, the self-dual $\bold{F}^+_{ab}$, and the scalar $C$ are descendant fields in the chiral multiplet. Up to notations, \eqref{chiral-polyform} coincides with \cite{BenettiGenolini:2026qdm}. By using \eqref{eq:gravitino-variation} and supersymmetry invariance of the chiral multiplet, $(\rd -\iota_\xi)\cJ^{{\rm chiral}}(\e)=0$ holds. A similar analysis follows for an anti-chiral action principle.
Note also that the $4d$ $\cN=2$ superspace action ($D$-term) emerges from the chiral/$F$-term action, see \cite{Freedman:2012zz,Lauria:2020rhc,Kuzenko:2022skv,Kuzenko:2022ajd}.

Another important action principle is given by the supersymmetric extension of a BF coupling \cite{deWit:1980gt,deWit:1982na,Lauria:2020rhc}. 
We show that this is also associated with an equivariantly closed polyform. The closed super 4-form for the BF action was constructed in \cite{Butter:2012ze}.
In the case without central charges, it is based on 
$J^{\rm BF} = \Sigma + V \wedge H$
where $\Sigma$ is constructed solely from covariant superfields and satisfies $\rd\Sigma=F\wedge H$, thereby making $J^{\rm BF}$ closed, $\rd J^{\rm BF}=0$. 
The associated polyform $\cJ^{\rm BF}(\e)=\e^{\iota_\e}J^{\rm BF}\tproj$ is
\bea
\cJ^{\rm BF}(\e)
&=& 
 \left(\frac{1}{8} \phi F + \frac{1}{8} \bar\phi \bar F + \frac{1}{32} X^{k l} G_{k l}\right){\rm vol}_4 
+v\wedge h
\non\\
&& 
+\frac{\ri}{2}(\phi\bar{\Upsilon}^{ij}
+\bar\phi\Upsilon^{ij}  ) G_{i j} 
-\frac{1}{2}v\wedge K^{ij} G_{i j}
\,,
\label{BF-polyform}
\eea
where
$
\Upsilon^{ij}
:=\frac{1}{2} e^{b} \wedge e^{a} (\e^{i} \s_{ab}\e^{j})
$.
When the supersymmetry conditions \eqref{eq:gravitino-variation}, \eqref{eq:gaugino-variation}, \eqref{eq:chi-variation}, and their conjugates, are satisfied, $(\rd-\iota_\xi)\cJ^{\rm BF}(\e)=0$ holds, provided that the supersymmetric gauge for the vector multiplet potential, 
%$
\begin{align}\label{eq:SupersymmetricGaugeCondition}
\xi^a v_a =( \bar{S}\phi- S\bar{\phi}) 
%$
 \,,
 \end{align}
is chosen. On supersymmetric backgrounds this condition ensures $\cL_\xi v=0$, see, e.g., \cite{BenettiGenolini:2024lbj}. In superspace, the analogous gauge choice $\cL_\e V=0$ implies $\cL_\e J^{\rm BF}=0$, and then \eqref{eq:dexpJ0b} and \eqref{eq:SuperspaceEquivariantForm} are satisfied \cite{note5}.

\section{Conclusion and outlook}
%\textit{Conclusion and outlook}---

We have shown that closed superforms, contracted with commuting Killing spinors, produce equivariantly closed polyforms. We then illustrated our proposal in off-shell $4d$ $\cN=2$ conformal supergravity.

It is important to determine more precisely how broadly our arguments apply.
A component-field analysis using gauge-covariant supergravity transformations rather than super-Lie derivatives should also be possible.
This may connect with the rheonomic approach to supergravity \cite{Castellani:1991eu}, where gravitino one-forms may be replaced by Killing spinors; see the discussion around Eq.~\eqref{Killing-spinor-expansion}. 

A next natural step is to systematically adapt our analysis to Euclidean supersymmetric backgrounds, where equivariant localization formulae are applied 
\cite{BenettiGenolini:2023kxp,Martelli:2023oqk,BenettiGenolini:2024xeo,Cassani:2024kjn,BenettiGenolini:2024lbj,Colombo:2025ihp,Colombo:2025yqy,Gaar:2026nqq,BenettiGenolini:2026hmz,BenettiGenolini:2026qdm,BenettiGenolini:2024kyy,BenettiGenolini:2025icr,Arav:2025jee,Park:2025fon,Arav:2026unc,Cassani:2026teb,Couzens:2026xmi,BenettiGenolini:2026cyc,BenettiGenolini:2026cdw}.
The construction should extend to broad classes of \(d\)-dimensional supergravities with Chern--Simons, Wess--Zumino, higher-form, and boundary couplings. In such cases, as already suggested by the BF example, gauge choices and boundary terms will play an important role \cite{Cassani:2024kjn,BenettiGenolini:2024lbj,Colombo:2025ihp,Colombo:2025yqy,Gaar:2026nqq,BenettiGenolini:2026hmz} and deserve a systematic superspace analysis.
For example, our superspace results, together with the framework of \cite{Howe:2011tm}, might be directly related to the relative equivariant cohomology results of \cite{BenettiGenolini:2026cdw}.

Using off-shell superconformal techniques in various dimensions, see, e.g., \cite{Freedman:2012zz,Lauria:2020rhc,Kuzenko:2022skv,Kuzenko:2022ajd}, one can adapt the present analysis to higher-derivative supergravities, see, e.g., \cite{Butter:2013lta,Butter:2014xxa,Butter:2013rba,Butter:2010jm,Kuzenko:2015jda,Ozkan:2024euj,Ma:2024ynp,Gold:2023ykx,Butter:2018wss,Hristov:2021qsw,Hristov:2022lcw,Hristov:2022plc,Hristov:2024cgj,BenettiGenolini:2026qdm,Hristov:2025ygn,Butter:2017jqu,Novak:2017wqc,Bobev:2020egg,Bobev:2021qxx,Bobev:2021oku,Cassani:2022lrk,Liu:2022sew,Gold:2023ymc}, with direct applications to holography. Superspace methods can also be employed for models with hypermultiplets \cite{Kuzenko:2008ep,Galperin:2001seg,Butter:2015nza}.

An important implication is a gravitino-counting criterion. From \eqref{gravitini-expansion} and \eqref{Killing-spinor-expansion}, the lowest-degree component of a supergravity polyform is dictated by the highest-gravitino term in its superform expansion. Up to exact-term subtleties, gauging data from Fayet--Iliopoulos terms or hypermultiplet moment maps \cite{Cremmer:1984hj,deWit:2001bk,Lauria:2020rhc,Kuzenko:2006nw,Butter:2014xua} enter through (composite) linear-multiplet BF couplings, which contain at most two gravitini and hence start at degree two; see \eqref{BF-polyform}. 
Hence, such terms should only contribute to bolts, in agreement with the absence of cosmological-constant/gauging terms from the nut formula of \cite{Hristov:2022lcw,Hristov:2024cgj}. In higher dimensions, this criterion may help identify fixed-component contributions and lead to non-renormalization theorems extending, e.g., \cite{Hristov:2025ygn}.

Looking further ahead, it would be worthwhile to revisit the superform structures of \(10d\) and \(11d\) supergravity; see, e.g., \cite{Castellani:1991eu,Cederwall:2001dx,Howe:2003cy,Berkovits:2008qw,Greitz:2011da,Soueres:2016qre,Grassi:2023yfe}. The aim would be to understand whether analogous equivariant constructions can be used to analyze BPS geometries, including compactified backgrounds. For example, combining the superspace results of \cite{Soueres:2016qre} with our general prescription for constructing polyforms leads to structures recently obtained for \(11d\) supergravity in \cite{BenettiGenolini:2026cyc,BenettiGenolini:2026cdw}. Further details will be reported elsewhere.

Our results indicate that the equivariant structures of supersymmetric gravitational observables are not accidental. They originate from superforms, with equivariant localization ingredients hidden in the fermionic directions of superspace and in the gravitino terms in supergravity. Thus, although the ``ectoplasm'' of supergravity may have no topology \cite{Gates:1997kr,Gates:1998hy}, it appears to encode substantial information about the topology of the bosonic body.

%\vspace{-5mm}
%%%%%%%%%%%%%%%%%
%\section*{Acknowledgements}
%%%%%%%%%%%%%%%%%
%\vspace{-3mm}
\begin{acknowledgments}
\medskip
\noindent\textbf{Acknowledgments}
%\textit{Acknowledgements.}---
G.T.-M. is grateful to S.\,J.\,Gates Jr.~for introducing him to superforms for the study of supersymmetric invariants and to D.\,Butter, S.\,M.\,Kuzenko, and J.\,Novak for years of collaboration on this subject. We are grateful to K.\,Hristov, S.\,Khandelwal, and Y.\,Pang for discussions and collaboration on related topics.
We are also grateful to S.\,J.\,Gates Jr., K.\,Hristov, S.\,M.\,Kuzenko, J.\,McOrist, and Y.\,Pang for comments on the manuscript.
This work has been supported by the Australian Research Council (ARC) Discovery Project DP240101409, the Capacity Building Package at the University of Queensland, and G.T.-M.'s faculty start-up funding from UQ’s School of Mathematics and Physics. 
C.K. and P.R. have been supported by postgraduate scholarships at the University of Queensland. 
G.T.-M. is also grateful to the organizers of the workshop ``Supergravity at 50'' at the Simons Center for Geometry and Physics, and to S.\,J.\,Gates Jr.\ at the University of Maryland, for their hospitality and support during visits made in the final stages of this work.

\end{acknowledgments}

%%%%%%%%%%%%%%%%%%%%%%%%%%%%%%%%%%%%%

\clearpage

\appendix{}

\onecolumngrid

\setcounter{section}{1}
 \setcounter{equation}{0}
 
\begin{center}\textbf{SUPPLEMENTAL MATERIAL\\(APPENDICES)}\end{center}

In the Supplemental Material to our Letter, we review useful general properties of superforms. We also collect the superform expressions used in our analysis of $4d$ $\cN=2$ conformal supergravity. These forms are defined on a $4d$ $\cN=2$ conformal superspace background, and the relevant results are taken from \cite{Butter:2012xg,Gold:2024gsj,Gates:2009xt,Butter:2017pbp,Butter:2012ze,Kuzenko:2022ajd}.

Given a rank-$p$ superform $\Omega$ with expansion
\bsubeq
\bea
\O
&=&
\frac{1}{p!}\rd z^{M_p}\wedge \cdots \wedge \rd z^{M_1}\O_{M_1\cdots M_p}
=\frac{1}{p!}E^{A_p}\wedge \cdots \wedge E^{A_1}\O_{A_1\cdots A_p}
~,
\eea
\esubeq
the wedge product of two superforms $\Omega$ and $\Omega^\prime$ satisfies
\begin{equation}
\Omega \wedge \Omega^\prime = (-1)^{|\Omega||\Omega^\prime| + \mathrm{rank}(\Omega) \mathrm{rank}(\Omega^\prime)} \Omega^\prime \wedge \Omega
\,.
\end{equation}
The exterior derivative ($\rd$) in superspace is 
\bsubeq\bea
\rd \O
&=&
\frac{1}{p!}\rd z^{M_p}\wedge\cdots \wedge\rd z^{M_1}\wedge\rd z^N\pa_{[N} \O_{M_1\cdots M_p)}
\,,
\\
&=&\frac{1}{p!}E^{A_p}\wedge\cdots \wedge E^{A_1}\wedge E^B\left(\de_{[B}\O_{A_1\cdots A_p)}
+\frac{p}{2}T_{[BA_1}{}^D\O_{|D|A_2\cdots A_p)}\right)
\,,
\eea
\esubeq
where in the second expression we have assumed $X_{\ua}\O=0$.

The interior product with a parity $|\mathscr X|$ vector field $\mathscr X$ is defined by $\iota_{{\mathscr X}} E^A ={\mathscr X}^A$ implying
\bsubeq
\bea
&\iota_{\mathscr X} \O
=
(-1)^{\mathrm{rank}(\Omega)-1}\dfrac{1}{(p-1)!}{\mathscr X}^{B} E^{A_{p-1}}\wedge\cdots \wedge E^{A_1} \O_{A_1\cdots A_{p-1} B}
~,\\
&\iota_{\mathscr X} \iota_{{\mathscr Y}}\Omega = -(-1)^{|{\mathscr X}||{\mathscr Y}|} \iota_{{\mathscr Y}} \iota_{\mathscr X}\Omega
\,.
\eea
\esubeq
The exterior derivative and interior product also satisfy
\bsubeq
\bea
\rd(\Omega\wedge \Omega^\prime) &=& \Omega \wedge \rd \Omega^\prime + (-1)^{{\rm rank}(\Omega^\prime)} \rd \Omega \wedge \Omega^\prime
\,,~~~~
\\
\iota_{\mathscr X}(\Omega
\wedge 
\Omega^\prime) 
&=& (-1)^{|{\mathscr X}||\Omega|} \Omega
\wedge 
\iota_{\mathscr X} \Omega^\prime
+ (-1)^{{\rm rank}(\Omega^\prime)} \iota_{\mathscr X} \Omega
\wedge 
\Omega^\prime
\,.~~~~
\eea
\esubeq

The vector multiplet closed super 2-form $F=F(W)$, $\rd F=0$, is
\bea
  F = \frac{1}{2} E^b \wedge E^a \ F_{a b} 
  -  \frac{\ri}{2} E^{\a}_{i} \wedge E^a \ (\s_{a})_{\a\bd} \bar\l^{\bd i} 
  + \frac{\ri}{2} \bar{E}^{i}_{\ad} \wedge E^a \  (\ts_{a})^{\ad\b} \l_{\b i} 
  - E^{i}_{\a} \wedge E^{\a}_{i} \ \bar W + \bar{E}^{\ad}_{i} \wedge \bar{E}^{i}_{\ad} \ W \, ,
\eea
where $W$ is a superconformal primary superfield of dimension 1 and $\sU(1)_{R}$ charge -2 satisfying 
\bsubeq
\bea
&\deb^{\ad}_iW=0~,~~~\de_\a^i\bar{W}=0
~,~~~\de^{ij}W=\bar{\de}^{ij}\bar{W}
\,,\\
&\l_\a^i=\nabla_{\a}^{i}W
~,~~~
\bar{\l}^\ad_i=\bar\nabla_{i}^{\ad}\bar W
\,,\\
& F_{a b} = -\dfrac{1}{8} (\s_{a b})_{\a \b}(\nabla^{\a \b}W + 4 W^{\a \b}\bar W) + \dfrac{1}{8}(\tilde \s_{a b})_{\ad \bd} (\bar\nabla^{\ad \bd} \bar W + 4 \bar W^{\ad \bd} W) \, ,
\eea
\esubeq 
where $\de^{ij}=\de^{\a(i}\de_\a^{j)}$, $\deb_{ij}= \deb_{\ad (i}\deb_{j)}^{\ad}$, $\de^{\a\b}:=\de^{(\a k}\de_k^{\b)}$, $\deb^{\ad\bd}=\deb_{k}^{(\ad}\deb^{\bd) k}$.

Provided the on-shell condition $\nabla^{i j}W =\bar{\de}^{i j}\bar{W} = X^{i j} = 0$ is satisfied, the vector multiplet not only admits  a real closed superform but also decomposes in terms of a complex closed super 2-form $\mathscr{P}$, and its conjugate $\bar{\mathscr{P}}$, $\rd\mathscr{P}=0$, $\rd\bar{\mathscr{P}}=0$ as
\begin{align}
   \mathscr{P}=\mathscr{P}(W) = \hf\big(F(W) + \ri F(\ri W)\big) \, , ~~~
   \mathscr{\bar P}=\mathscr{\bar P}(W) = \hf\big(F(W) - \ri F(\ri W)\big)\,,~~~
F(W)= \mathscr{P}+\mathscr{\bar P}
\ .
\end{align}
The explicit expansions of these superforms are
\bsubeq
\begin{align}
    \mathscr{P} &= \frac{1}{2} E^{b} \wedge E^{a} \, \mathscr{P}_{a b} 
    - \frac{\ri}{2} E^{\a}_{i} \wedge E^{a} \,  (\s_{a})_{\a\bd} \bar\l^{\bd i} 
    - E^{\a}_{i} \wedge E_{\a}^{i} \, \bar W \, , \\
    \mathscr{\bar P} &= \frac{1}{2} E^{b} \wedge E^{a} \, \mathscr{\bar P}_{a b} 
    + \frac{\ri}{2} \bar{E}_{\ad}^{i} \wedge E^{a} \, (\tilde{\s}_{a})^{\ad\b} \l_{\b i}
    + \bar{E}_{\ad}^{i} \wedge \bar{E}_{\ad}^{i} \, W \, , 
\end{align}
\esubeq
where  
\bea
\mathscr{P}_{a b} = -\frac{1}{2} W_{a b}^{+} \bar W + \frac{1}{8} (\tilde \s_{a b})_{\ad \bd} \bar \nabla^{\ad \bd} \bar W \,,~~~~~~\mathscr{\bar P}_{ab} = - \frac{1}{2} W_{ab}^{-} W - \frac{1}{8} (\s_{a b})_{\a \b}\nabla^{\a \b} W\,.
\eea

The linear multiplet closed super 3-form $H=H(G)$, $\rd H=0$, is
\bea
    H &=& \frac{1}{6} E^{c} \wedge E^{b} \wedge E^{a} \ H_{abc} + \frac{\ri}{4} E^{\a}_{i} \wedge E^{b} \wedge E^{a} \ (\s_{a b})_{\a}{}^{\b} \chi_{\b}^{i}
    + \frac{\ri}{4} \bar{E}_{\ad}^{i} \wedge E^{b} \wedge E^{a} \ (\tilde \s_{a b})^{\ad}{}_{\bd} \bar\chi^{\bd}_{i}
    \non \\
    && + \frac{1}{2} \bar{E}_{\ad}^{i} \wedge E_{i}^{\a} \wedge E^{a} \ (\s_{a})_{\a}{}^{\ad} G^{i}{}_{j}
    \,,
\eea
where $G^{i j}$ is a real, dimension 2 symmetric $G^{i j}=G^{ji}$ primary superfield satisfying
\bea
&\de_{\a}^{(i}G^{jk)}=\bar{\de}_{\ad}^{(i}G^{jk)}=0
\,,\\
&\chi_{\a i}=\dfrac{1}{3} \nabla_{\a}^{j}G_{i j}\,,~~~
\bar{\chi}^{\ad i}= \dfrac{1}{3} \bar\nabla^{\ad}_{j} G^{i j}
\,,\\
    &H_{a b c} = \dfrac{\ri}{96} \ve_{abcd} (\s^{d})^{\a}{}_{\bd} [\nabla_{\a}^{i}, \bar\nabla_{j}^{\bd}] G^{j}{}_{i}
    \,.
\eea

Given a superconformal primary superfield $\cL_c$ of dimension 2 and $\sU(1)_{R}$ charge -4 satisfying 
\bea
\deb^\ad_i\cL_c=0\,,
\eea 
 but otherwise arbitrary, the closed 4-form $J^{\text{chiral}}$ which describes the chiral action principle \cite{Gates:2009xt,Kuzenko:2022ajd} is
\bea
J^{\text{chiral}} &=& 4\ri \bar{E}_{\bd}^{j} \wedge \bar{E}_{j}^{\bd} \wedge \bar{E}_{\ad}^{i} \wedge \bar{E}_{i}^{\ad} \ \cL_{c} 
+ 2 \bar{E}_{\bd}^{j} \wedge \bar{E}_{j}^{\bd} \wedge \bar{E}_{\ad}^{i} \wedge E^{a} \ (\tilde \s_{a})^{\ad \a} \Psi_{\a i}
\non \\
& &
- \ri\bar{E}_{\bd}^{j} \wedge \bar{E}_{\ad}^{i} \wedge E^{b} \wedge E^{a} \ (\tilde \s_{a b})^{\ad \bd} B_{i j} 
- \ri \bar{E}_{\ad}^{i} \wedge \bar{E}^{\ad}_{i} \wedge E^{b} \wedge E^{a} \left((\s_{ab})_{\a \b} \bold{F}^{\a \b} - 2(\tilde \s_{a b})_{\ad \bd}\bar W^{\ad \bd}\cL_{c} \right) 
\non \\
        & &
        - \frac{\ri}{6}\ve_{abcd} \bar{E}_{\ad}^{i} \wedge E^{c} \wedge E^{b} \wedge E^{a} \left( (\tilde \s^{d})^{\ad\a}\L_{\a i}   -  (\s^{d})_{\a\bd} \bar W^{\ad \bd} \Psi^\a_{i}\right)
        \non \\
        & &
+ \frac{1}{24} \ve_{abcd} E^{d} \wedge E^{c} \wedge E^{b} \wedge E^{a} \left(\hf C + \bar W^{\ad \bd} \bar W_{\ad \bd}\cL_{c} \right)  \, ,
\eea
where $\bar{W}^{\ad\bd}=-\hf (\tilde{\s}^{ab})^{\ad\bd} W_{ab}$, and
\bea
\Psi_{\a i} = \nabla_{\a i}\cL_{c}
\,,~~~ 
B_{i j} = \hf \nabla_{i j}\cL_c\,,~~~ \bold{F}^{\a \b} = \frac{1}{4}\nabla^{\a \b}\cL_{c}
\,,~~~
\L_{\a i} = \frac{1}{6} \nabla_{\a}^{j}\nabla_{i j}\cL_{c}
\,, ~~~C = \frac{1}{24}\de^{ij}\de_{ij}\cL_{c}
\,.
\eea 
In the Letter we used $\Phi$ for $\cL_c$. 
As an interesting observation, note that in \cite{Gates:2009xt} it was shown that the wedge product of two closed complex on-shell vector multiplet super 2-forms, $\bar{\mathscr P} \wedge \bar{\mathscr P}$, coincides with $J^{\text{chiral}}$ when choosing $\cL_{c} = -\frac{\ri}{4} W^2$. This can also be shown to hold for the associated polyforms presented in our Letter.
One has
$\cJ^{\rm chiral}(\e)\big|_{\Phi= -\frac{\ri}{4} \phi^2} 
= \bar\cG(\e) \wedge \bar\cG(\e)$
when $\Phi=-\frac{\ri}{4}\phi^2$ as well as the condition $X^{ij}=0$, and the descendants 
$B^{ij}= 0$, 
$\bold{F}^+_{a b} = \ri \bar g_{a b}\phi + \frac{\ri}{2}  W_{ab}^{-} \phi^2$, 
together with 
$C 
= 
\ri\big(
\bar g_{ab} \bar g^{ab} 
+W_{a b}^{-} \bar g^{a b} \phi 
+\frac{1}{4} W_{a b}^{-}{W^{-}}^{a b}\phi^2 \big)$. 
This is with $f_{a b}^{+} = \bar g_{a b} + \frac{1}{2} W_{a b}^{-} \phi - \frac{1}{2} W_{a b}^{+}\bar \phi$.

\vspace{0.3cm}
The closed 4-form $J^{\text{BF}} = \S + V \wedge H$ describes the BF action. The super 4-form $\S$ is
\bea
\Sigma
&=& \frac{1}{24} E^{d} \wedge E^{c} \wedge E^{b} \wedge E^{a} \ \ve_{abcd}\left(\frac{1}{8} WF + \frac{1}{8} \bar W \bar F + \frac{1}{32} X^{k l} G_{k l} - \frac{1}{8} \l^{\g}_{k} \chi_{\g}^{ k} 
- \frac{1}{8} \bar\l_{\gd}^{ k}\bar\chi^{\gd}_{ k}\right)
\non\\
        &&
        + \frac{1}{6} E_{i}^{\a} \wedge E^{c} \wedge E^{b} \wedge E^{a} \ \left(- \frac{\ri}{8} \varepsilon_{abcd} (\s^{d})_{\a \ad} \bar \l_{j}^{\ad}  G^{i j} - \frac{\ri}{4} \varepsilon_{abcd}(\s^{d})_{\a\ad} \bar W \bar \chi^{\ad i}\right)
        \non\\
        && 
        + \frac{1}{6} \bar{E}^{i}_{\ad} \wedge E^{c} \wedge E^{b} \wedge E^{a} \ \left(
       - \frac{\ri}{8}\ve_{abcd} (\tilde{\s}^{d})^{\ad\a} \l^{j}_{\a} G_{i j} 
       -\frac{\ri}{4} \ve_{abcd} (\tilde{\s}^{d})^{\ad\a} W \chi_{\a i}\right)   \non\\
        && 
        + \frac{\ri}{4} E_{j}^{\b} \wedge E_{i}^{\a} \wedge E^{b} \wedge E^{a} \ (\s_{a b})_{\a \b} \bar W G^{i j} 
        - \frac{\ri}{4} \bar{E}_{j}^{\bd} \wedge \bar{E}_{i}^{\ad} \wedge E^{b} \wedge E^{a} \ (\tilde \s_{a b})_{\ad \bd} W G^{i j}
        \,,
    \eea
with  $F=\dfrac{1}{12}\de^{ij}G_{ij}$ 
and 
$\bar{F}=\dfrac{1}{12}\deb^{ij}G_{ij}$.
The expression for $V\wedge H$ is quite lengthy. For our purposes, however, it suffices to note that the vector multiplet potential super 1-form $V=E^aV_a+E^\a_i V_\a^i+ \bar{E}_\ad^i \bar{V}^\ad_i$ is constrained by the Wess-Zumino-type gauge condition $V^{\a}_i\loco=\bar{V}_\ad^i\loco=0$, implying $V\doubar= e^a v_a$, which simplifies $V\wedge H\doubar=v\wedge h$.


\begin{thebibliography}{2}


\bibitem{Pestun:2016zxk}
V.~Pestun, M.~Zabzine, F.~Benini, T.~Dimofte, T.~T.~Dumitrescu, K.~Hosomichi, S.~Kim, K.~Lee, B.~Le Floch and M.~Marino, \textit{et al.}
\doihref{10.1088/1751-8121/aa63c1}
{J. Phys. A \textbf{50}, no.44, 440301 (2017)},
\arxivlink{1608.02952}.


\bibitem{Atiyah:1984px}
M.~F.~Atiyah and R.~Bott,
\doihref{10.1016/0040-9383(84)90021-1}
{Topology \textbf{23} (1984), 1-28}.


\bibitem{Berline:1982}
N.~Berline and M.~Vergne,
%``Classes caract\'eristiques \'equivariantes. Formule de localisation en cohomologie \'equivariante,''
C. R. Acad. Sci. Paris S\'er. I Math. \textbf{295} (1982) no.~9, 539--541.


\bibitem{BenettiGenolini:2023kxp}
P.~Benetti Genolini, J.~P.~Gauntlett and J.~Sparks,
\doihref{10.1103/PhysRevLett.131.121602}
{Phys. Rev. Lett. \textbf{131} (2023) no.12, 121602}.
%\arxivlink{2306.03868}.


\bibitem{Martelli:2023oqk}
D.~Martelli and A.~Zaffaroni,
\doihref{10.1007/s11005-023-01752-1}
{Lett. Math. Phys. \textbf{114} (2024) no.1, 15}.
%\arxivlink{2306.03891}.


\bibitem{Dabholkar:2010uh}
A.~Dabholkar, J.~Gomes and S.~Murthy,
\doihref{10.1007/JHEP06(2011)019}
{JHEP \textbf{06} (2011), 019}.
%\arxivlink{1012.0265}.


\bibitem{Hristov:2018lod}
K.~Hristov, I.~Lodato and V.~Reys,
\doihref{10.1007/JHEP07(2018)072}
{JHEP \textbf{07} (2018), 072}.
%\arxivlink{1803.05920}.


\bibitem{Hosseini:2019iad}
S.~M.~Hosseini, K.~Hristov and A.~Zaffaroni,
\doihref{10.1007/JHEP12(2019)168}
{JHEP \textbf{12} (2019), 168}.
%\arxivlink{1909.10550}.




\bibitem{Hristov:2021qsw}
K.~Hristov,
\doihref{10.1007/JHEP02(2022)079}
{JHEP \textbf{02} (2022), 079}.
%\arxivlink{2111.06903}.


\bibitem{Hristov:2022lcw}
K.~Hristov,
\doihref{10.1007/JHEP10(2022)190}
{JHEP \textbf{10} (2022), 190}.
%\arxivlink{2204.02992}.


\bibitem{Hristov:2022plc}
K.~Hristov,
%``Maximally symmetric nuts in 4d {\ensuremath{\mathscr{N}}} = 2 higher derivative supergravity,''
\doihref{10.1007/JHEP02(2023)110}
{JHEP \textbf{02}, 110 (2023)}.
%\arxivlink{2212.10590}.

\bibitem{Colombo:2023fhu}
E.~Colombo, F.~Faedo, D.~Martelli and A.~Zaffaroni,
\doihref{10.1007/JHEP01(2024)095}
{JHEP \textbf{01} (2024), 095}.
%\arxivlink{2309.04425}.



\bibitem{BenettiGenolini:2024kyy}
P.~Benetti Genolini, J.~P.~Gauntlett, Y.~Jiao, A.~L{\"u}scher and J.~Sparks,
%``Localization and attraction,''
\doihref{10.1007/JHEP05(2024)152}
{JHEP \textbf{05} (2024), 152}.
%[arXiv:2401.10977 [hep-th]].

\bibitem{Hristov:2024cgj}
K.~Hristov, 
\emph{Equivariant localization and gluing rules in 4d N = 2 higher derivative supergravity}, in: \doihref{10.1007/978-3-032-16202-1_4}{Wood, D.R., Etheridge, A.M., de Gier, J., Joshi, N. (eds) 2024 MATRIX Annals, Part I. MATRIX Book Series, vol 7. Springer, Cham}. 
%\arxivlink{2406.18648}.


\bibitem{BenettiGenolini:2024xeo}
P.~Benetti Genolini, J.~P.~Gauntlett, Y.~Jiao, A.~L{\"u}scher and J.~Sparks,
\doihref{10.1103/PhysRevLett.133.141601}
{Phys. Rev. Lett. \textbf{133} (2024) no.14, 141601}.
%\arxivlink{2407.02554}.


\bibitem{Cassani:2024kjn}
D.~Cassani, A.~Ruip{\'e}rez and E.~Turetta,
\doihref{10.1007/JHEP12(2024)086}
{JHEP \textbf{12}, 086 (2024)}.
%\arxivlink{2409.01332}.


\bibitem{BenettiGenolini:2024lbj}
P.~Benetti Genolini, J.~P.~Gauntlett, Y.~Jiao, A.~L{\"u}scher and J.~Sparks,
\doihref{10.1007/JHEP08(2025)211}
{JHEP \textbf{08} (2025), 211}.
%\arxivlink{2412.07828}.


\bibitem{Colombo:2025ihp}
E.~Colombo, V.~Dimitrov, D.~Martelli and A.~Zaffaroni,
\arxivlink{2502.15624}.




\bibitem{BenettiGenolini:2025icr}
P.~Benetti Genolini, J.~P.~Gauntlett, Y.~Jiao, J.~Park and J.~Sparks,
%``Equivariant localization for D = 5 gauged supergravity,''
\doihref{10.1007/JHEP03(2026)080}
{JHEP \textbf{03} (2026), 080}.
%[arXiv:2508.08207 [hep-th]].




\bibitem{Arav:2025jee}
I.~Arav, J.~P.~Gauntlett, M.~M.~Roberts and C.~Rosen,
%``Spindle solutions, hyperscalars and smooth uplifts,''
\doihref{10.1007/JHEP06(2026)010}
{JHEP \textbf{06} (2026), 010}.
%[arXiv:2511.01964 [hep-th]].

\bibitem{Colombo:2025yqy}
E.~Colombo, V.~Dimitrov, D.~Martelli and A.~Zaffaroni,
\arxivlink{2511.13824}.



\bibitem{Park:2025fon}
J.~Park,
%``Localizing AlAdS$_{5}$ black holes and the SUSY index on S$^{1}${\texttimes} M$_{3}$,''
\doihref{10.1007/JHEP06(2026)107}
{JHEP \textbf{06} (2026), 107}.
%[arXiv:2511.15666 [hep-th]].


\bibitem{Gaar:2026nqq}
F.~Gaar, J.~P.~Gauntlett, J.~Park and J.~Sparks,
\arxivlink{2604.09490}.


\bibitem{BenettiGenolini:2026hmz}
P.~Benetti Genolini, C.~Couzens and A.~L{\"u}scher,
\arxivlink{2604.26786}.


\bibitem{BenettiGenolini:2026qdm}
P.~Benetti Genolini, F.~Gaar, J.~P.~Gauntlett and J.~Sparks,
\arxivlink{2604.08656}.


\bibitem{Arav:2026unc}
I.~Arav, J.~P.~Gauntlett, J.~Park, M.~M.~Roberts and C.~Rosen,
%``Spindle solutions with hyperscalars in $D=4$ gauged supergravity,''
\arxivlink{2605.04140}.

\bibitem{Cassani:2026teb}
D.~Cassani and E.~Turetta,
%``The black hole at the end of the cone: localizing the anomaly polynomial on toric geometries,''
\arxivlink{2606.16986}.

\bibitem{Couzens:2026xmi}
C.~Couzens, A.~L{\"u}scher and J.~Sparks,
%``Ten-dimensional localization,''
\arxivlink{2607.05529}.

\bibitem{BenettiGenolini:2026cyc}
P.~Benetti Genolini, F.~Gaar, J.~P.~Gauntlett, J.~Park and J.~Sparks,
%``Airy functions from quantum M-theory,''
\arxivlink{2607.07255}.

\bibitem{BenettiGenolini:2026cdw}
P.~Benetti Genolini, F.~Gaar, J.~P.~Gauntlett, J.~Park and J.~Sparks,
%``Odd-Dimensional Localization in Supergravity,''
\arxivlink{2607.07621}.











\bibitem{Castellani:1991eu}
L.~Castellani, R.~D'Auria and P.~Fre,
\doihref{10.1142/0224}
{\emph{Supergravity and superstrings: A Geometric perspective. Vol. 2: Supergravity}},
World Scientific, Singapore, 1991,
pp.~680--684.


\bibitem{Gates:1997kr}
S.~J.~Gates, Jr.,
\arxivlink{hep-th/9709104}.


\bibitem{Gates:1997ag}
S.~J.~Gates, Jr., M.~T.~Grisaru, M.~E.~Knutt-Wehlau and W.~Siegel,
\doihref{10.1016/S0370-2693(97)01557-8}
{Phys. Lett. B \textbf{421}, 203-210 (1998)}.
%\arxivlink{hep-th/9711151}.


\bibitem{Gates:1998hy}
S.~J.~Gates, Jr.,
\doihref{10.1016/S0550-3213(98)00819-0}
{Nucl. Phys. B \textbf{541}, 615-650 (1999)}.
%\arxivlink{hep-th/9809056}.



\bibitem{Howe:2003cy}
P.~S.~Howe and D.~Tsimpis,
\doihref{10.1088/1126-6708/2003/09/038}
{JHEP \textbf{09}, 038 (2003)}.
%\arxivlink{hep-th/0305129}.


\bibitem{Berkovits:2006ik}
N.~Berkovits,
\arxivlink{hep-th/0612021}.


\bibitem{Berkovits:2008qw}
N.~Berkovits and P.~S.~Howe,
\doihref{10.1088/1126-6708/2008/06/046}
{JHEP \textbf{06}, 046 (2008)}.
%\arxivlink{0803.3024}.


\bibitem{Gates:2009uv}
S.~J.~Gates, Jr. and G.~Tartaglino-Mazzucchelli,
\doihref{10.1088/1751-8113/43/9/095401}
{J. Phys. A \textbf{43}, 095401 (2010)}.
%\arxivlink{0907.5264}.


\bibitem{Gates:2009xt}
S.~J.~Gates, Jr., S.~M.~Kuzenko and G.~Tartaglino-Mazzucchelli,
\doihref{10.1103/PhysRevD.80.125015}
{Phys. Rev. D \textbf{80}, 125015 (2009)}.
%\arxivlink{0909.3918}.

\bibitem{Howe:2011tm}
P.~S.~Howe, T.~G.~Pugh, K.~S.~Stelle and C.~Strickland-Constable,
%``Ectoplasm with an Edge,''
\doihref{10.1007/JHEP08(2011)081}
{JHEP \textbf{08} (2011), 081}.
%[arXiv:1104.4387 [hep-th]].

\bibitem{Butter:2012ze}
D.~Butter, S.~M.~Kuzenko and J.~Novak,
\doihref{10.1007/JHEP09(2012)131}
{JHEP \textbf{09}, 131 (2012)}.
%\arxivlink{1205.6981}.


\bibitem{Kuzenko:2012ew}
S.~M.~Kuzenko and G.~Tartaglino-Mazzucchelli,
\doihref{10.1007/JHEP03(2013)113}
{JHEP \textbf{03}, 113 (2013)}.
%\arxivlink{1212.6852}.


\bibitem{Butter:2013rba}
D.~Butter, S.~M.~Kuzenko, J.~Novak and G.~Tartaglino-Mazzucchelli,
\doihref{10.1007/JHEP10(2013)073}
{JHEP \textbf{10}, 073 (2013)}.
%\arxivlink{1306.1205}.


\bibitem{Kuzenko:2013rna}
S.~M.~Kuzenko and J.~Novak,
\doihref{10.1007/JHEP02(2014)096}
{JHEP \textbf{02}, 096 (2014)}.
%\arxivlink{1309.6803}.


\bibitem{Kuzenko:2014jra}
S.~M.~Kuzenko and J.~Novak,
\doihref{10.1007/JHEP05(2014)093}
{JHEP \textbf{05}, 093 (2014)}.
%\arxivlink{1401.2307}.


\bibitem{Arias:2014ona}
C.~Arias, W.~D.~Linch, III and A.~K.~Ridgway,
\doihref{10.1007/JHEP05(2016)016}
{JHEP \textbf{05}, 016 (2016)}.
%\arxivlink{1402.4823}.


\bibitem{Castellani:2014goa}
L.~Castellani, R.~Catenacci and P.~A.~Grassi,
\doihref{10.1016/j.nuclphysb.2014.10.023}
{Nucl. Phys. B \textbf{889}, 419-442 (2014)}.
%\arxivlink{1409.0192}.


\bibitem{Gates:2014cqa}
S.~J.~Gates, W.~D.~Linch and S.~Randall,
\doihref{10.1007/JHEP05(2015)049}
{JHEP \textbf{05}, 049 (2015)}.
%\arxivlink{1412.4086}.


\bibitem{Linch:2014iza}
W.~D.~Linch and S.~Randall,
\doihref{10.1007/JHEP09(2015)190}
{JHEP \textbf{09}, 190 (2015)}.
%\arxivlink{1412.4686}.


\bibitem{Castellani:2015paa}
L.~Castellani, R.~Catenacci and P.~A.~Grassi,
\doihref{10.1016/j.nuclphysb.2015.07.028}
{Nucl. Phys. B \textbf{899}, 112-148 (2015)}.
%\arxivlink{1503.07886}.


\bibitem{Butter:2016qkx}
D.~Butter, S.~M.~Kuzenko, J.~Novak and S.~Theisen,
\doihref{10.1007/JHEP12(2016)072}
{JHEP \textbf{12}, 072 (2016)}.
%\arxivlink{1606.02921}.


\bibitem{Castellani:2016ibp}
L.~Castellani, R.~Catenacci and P.~A.~Grassi,
\doihref{10.1007/JHEP10(2016)049}
{JHEP \textbf{10}, 049 (2016)}.
%\arxivlink{1607.05193}.


\bibitem{Soueres:2016qre}
B.~Sou{\`e}res and D.~Tsimpis,
%``Action principle and the supersymmetrization of Chern-Simons terms in eleven-dimensional supergravity,''
\doihref{10.1103/PhysRevD.95.026013}
{Phys. Rev. D \textbf{95} (2017) no.2, 026013}.
%[arXiv:1612.02021 [hep-th]].


\bibitem{Becker:2017njd}
K.~Becker, M.~Becker, W.~D.~Linch, III, S.~Randall and D.~Robbins,
\doihref{10.1007/JHEP04(2017)103}
{JHEP \textbf{04}, 103 (2017)}.
%\arxivlink{1702.00799}.


\bibitem{Butter:2018wss}
D.~Butter, J.~Novak, M.~Ozkan, Y.~Pang and G.~Tartaglino-Mazzucchelli,
\doihref{10.1007/JHEP04(2019)013}
{JHEP \textbf{04} (2019), 013}.
%\arxivlink{1808.00459}.


\bibitem{Butter:2019edc}
D.~Butter, F.~Ciceri and B.~Sahoo,
\doihref{10.1007/JHEP01(2020)029}
{JHEP \textbf{01}, 029 (2020)}.
%\arxivlink{1910.11874}.


\bibitem{Cremonini:2021vyy}
C.~A.~Cremonini and P.~A.~Grassi,
\arxivlink{2105.09196}.


\bibitem{Grassi:2023bxm}
P.~A.~Grassi,
\doihref{10.3390/universe9080376}
{Universe \textbf{9}, no.8, 376 (2023)}.
%\arxivlink{2304.04877}.


\bibitem{Grassi:2023yfe}
P.~A.~Grassi,
\doihref{10.1088/1742-6596/2531/1/012010}
{J. Phys. Conf. Ser. \textbf{2531}, no.1, 012010 (2023)}.
%\arxivlink{2304.01743}.







\bibitem{SUPERSPACE}
S.~J.~Gates, M.~T.~Grisaru, M.~Rocek and W.~Siegel,
\href{https://arxiv.org/abs/hep-th/0108200}
{\emph{Superspace Or One Thousand and One Lessons in Supersymmetry},
Front. Phys. \textbf{58}, 1--548 (1983)}.


\bibitem{Wess:1992cp}
J.~Wess and J.~Bagger,
\href{https://inspirehep.net/literature/350988}
{\emph{Supersymmetry and supergravity},
Princeton University Press, 1992}.



\bibitem{Buchbinder-Kuzenko}
I.~Buchbinder and S.~M.~Kuzenko.
\href{https://inspirehep.net/literature/1946734}{Ideas and methods of supersymmetry and supergravity:
Or a walk  through superspace,
IOP, Bristol (1998)}.


\bibitem{DeWitt:2012mdz}
B.~S.~DeWitt,
\emph{Supermanifolds},
\doihref{10.1017/CBO9780511564000}
{Cambridge Univ. Press, 2012}.


\bibitem{Kuzenko:2022skv}
S.~M.~Kuzenko, E.~S.~N.~Raptakis and G.~Tartaglino-Mazzucchelli,
\doihref{10.1007/978-981-19-3079-9_40-1}{\emph{Superspace Approaches to  $\cN=1$ Supergravity}, in: Bambi, C., Modesto, L., Shapiro, I. (eds) Handbook of Quantum Gravity. Springer, Singapore},
\arxivlink{2210.17088}.


\bibitem{Kuzenko:2022ajd}
S.~M.~Kuzenko, E.~S.~N.~Raptakis and G.~Tartaglino-Mazzucchelli,
\doihref{10.1007/978-981-19-3079-9_44-1}{\emph{Covariant Superspace Approaches to $\cN=2$ Supergravity}, in: Bambi, C., Modesto, L., Shapiro, I. (eds) Handbook of Quantum Gravity. Springer, Singapore},
\arxivlink{2211.11162}.


\bibitem{Gates:1980ay}
S.~J.~Gates, Jr.,
\doihref{10.1016/0550-3213(81)90225-X}
{Nucl. Phys. B \textbf{184}, 381-390 (1981)}.


\bibitem{Bergshoeff:1996tu}
E.~Bergshoeff and P.~K.~Townsend,
%``Super D-branes,''
\doihref{10.1016/S0550-3213(97)00072-2}{Nucl. Phys. B \textbf{490}, 145-162 (1997)}.
% doi:10.1016/S0550-3213(97)00072-2
% [arXiv:hep-th/9611173 [hep-th]].


\bibitem{Bandos:1997ui}
I.~A.~Bandos, K.~Lechner, A.~Nurmagambetov, P.~Pasti, D.~P.~Sorokin and M.~Tonin,
%``Covariant action for the superfive-brane of M theory,''
\doihref{10.1103/PhysRevLett.78.4332}{Phys. Rev. Lett. \textbf{78}, 4332-4334 (1997)}.
%doi:10.1103/PhysRevLett.78.4332
%[arXiv:hep-th/9701149 [hep-th]].

\bibitem{Cederwall:2001dx}
M.~Cederwall, B.~E.~W.~Nilsson and D.~Tsimpis,
\doihref{10.1088/1126-6708/2002/02/009}
{JHEP \textbf{02}, 009 (2002)}.
%\arxivlink{hep-th/0110069}.

\bibitem{Greitz:2011da}
J.~Greitz and P.~S.~Howe,
\doihref{10.1007/JHEP08(2011)146}
{JHEP \textbf{08}, 146 (2011)}.
%\arxivlink{1103.5053}.

\bibitem{Kennedy:2025nzm}
C.~Kennedy and G.~Tartaglino-Mazzucchelli,
\doihref{10.1007/JHEP08(2025)215}
{JHEP \textbf{08}, 215 (2025)}.
%\arxivlink{2506.01630}.



\bibitem{note1}
Or, transforming into an exact form, $\d_\L J=\rd j_\L$.

\bibitem{note2}
Related ideas appear in the mathematical and physical literature on equivariant localization; see, e.g., \cite{Pestun:2016zxk}. To the best of our knowledge, however, the general mechanism presented here relating closed superforms to equivariantly closed polyforms is new in the supergravity literature, and constitutes the main result of this Letter.




\bibitem{deWit:1979dzm}
B.~de Wit, J.~W.~van Holten and A.~Van Proeyen,
\doihref{10.1016/0550-3213(80)90125-X}
{Nucl. Phys. B \textbf{167}, 186 (1980)}.


\bibitem{deWit:1980lyi}
B.~de Wit, J.~W.~van Holten and A.~Van Proeyen,
\doihref{10.1016/0550-3213(83)90548-5}
{Nucl. Phys. B \textbf{184}, 77 (1981) [erratum: Nucl. Phys. B \textbf{222}, 516 (1983)]}.


\bibitem{deWit:1980gt}
B.~de Wit, J.~W.~van Holten and A.~Van Proeyen,
\doihref{10.1016/0370-2693(80)90397-4}
{Phys. Lett. B \textbf{95}, 51-55 (1980)}.


\bibitem{deWit:1982na}
B.~de Wit, R.~Philippe and A.~Van Proeyen,
\doihref{10.1016/0550-3213(83)90432-7}
{Nucl. Phys. B \textbf{219}, 143-166 (1983)}.


\bibitem{deWit:1984wbb}
B.~de Wit and A.~Van Proeyen,
\doihref{10.1016/0550-3213(84)90425-5}
{Nucl. Phys. B \textbf{245}, 89-117 (1984)}.


\bibitem{deWit:1984rvr}
B.~de Wit, P.~G.~Lauwers and A.~Van Proeyen,
\doihref{10.1016/0550-3213(85)90154-3}
{Nucl. Phys. B \textbf{255}, 569-608 (1985)}.


\bibitem{Freedman:2012zz}
D.~Z.~Freedman and A.~Van Proeyen,
\doihref{10.1017/CBO9781139026833}
{\emph{Supergravity},
Cambridge Univ. Press, 2012}.

\bibitem{Lauria:2020rhc}
E. Lauria and A. Van Proeyen,
\emph{${\cal N}=2$ Supergravity in $D=4,5,6$ Dimensions},
\doihref{10.1007/978-3-030-33757-5}{Lect. Notes Phys. \textbf{966}, 3, 2020}.


\bibitem{Butter:2009cp}
D.~Butter,
%``N=1 Conformal Superspace in Four Dimensions,''
\doihref{10.1016/j.aop.2009.09.010}
{Annals Phys. \textbf{325} (2010), 1026-1080}.


\bibitem{Butter:2011sr}
D.~Butter,
\doihref{10.1007/JHEP10(2011)030}
{JHEP \textbf{10}, 030 (2011)}.
%\arxivlink{1103.5914}.


\bibitem{Butter:2012xg}
D.~Butter and J.~Novak,
\doihref{10.1007/JHEP05(2012)115}
{JHEP \textbf{05}, 115 (2012)}.
%\arxivlink{1201.5431}.


\bibitem{Gold:2024gsj}
G.~Gold, S.~Khandelwal and G.~Tartaglino-Mazzucchelli,
\doihref{10.1007/JHEP02(2025)196}
{JHEP \textbf{02}, 196 (2025)}.
%\arxivlink{2409.19034}.


\bibitem{Klare:2012gn}
C.~Klare, A.~Tomasiello and A.~Zaffaroni,
\doihref{10.1007/JHEP08(2012)061}
{JHEP \textbf{08}, 061 (2012)}.
%\arxivlink{1205.1062}.


\bibitem{Cassani:2012ri}
D.~Cassani, C.~Klare, D.~Martelli, A.~Tomasiello and A.~Zaffaroni,
\doihref{10.1007/s00220-014-1983-3}
{Commun. Math. Phys. \textbf{327}, 577-602 (2014)}.
%\arxivlink{1207.2181}.


\bibitem{Klare:2013dka}
C.~Klare and A.~Zaffaroni,
\doihref{10.1007/JHEP10(2013)218}
{JHEP \textbf{10}, 218 (2013)}.
%\arxivlink{1308.1102}.



\bibitem{Kuzenko:2015lca}
S.~M.~Kuzenko,
\doihref{10.22323/1.231.0140}
{PoS \textbf{CORFU2014} (2015), 140}.
%\arxivlink{1504.08114}.



\bibitem{note4}
This assumption can be relaxed.


\bibitem{Butter:2015tra}
D.~Butter, G.~Inverso and I.~Lodato,
\doihref{10.1007/JHEP09(2015)088}
{JHEP \textbf{09} (2015), 088}.
%\arxivlink{1505.03500}.


\bibitem{deWit:2017cle}
B.~de Wit and V.~Reys,
%``Euclidean supergravity,''
\doihref{10.1007/JHEP12(2017)011}{JHEP \textbf{12} (2017), 011}.
%[arXiv:1706.04973 [hep-th]].


\bibitem{note3}
The on-shell vector multiplet also plays a central role in the construction of the variant \(4d\) \(\cN=2\) dilaton-Weyl multiplet of conformal supergravity \cite{Butter:2017pbp}.




\bibitem{Butter:2017pbp}
D.~Butter, S.~Hegde, I.~Lodato and B.~Sahoo,
\doihref{10.1007/JHEP03(2018)154}
{JHEP \textbf{03}, 154 (2018)}.
%\arxivlink{1712.05365}.


\bibitem{note5}
The gauge condition $\cL_\e V=0$ can be extended in superspace to ensure \eqref{eq:SupersymmetricGaugeCondition} is shifted by constants, allowing regularity of $v_a$ when $\xi^a=0$ as, for example, in \cite{BenettiGenolini:2024lbj}. This extension and associated modifications to $\cJ^{\rm BF}(\e)$ will be presented elsewhere.



\bibitem{Butter:2010jm}
D.~Butter and S.~M.~Kuzenko,
\doihref{10.1007/JHEP03(2011)047}
{JHEP \textbf{03} (2011), 047}.
%\arxivlink{1012.5153}.


\bibitem{Butter:2013lta}
D.~Butter, B.~de Wit, S.~M.~Kuzenko and I.~Lodato,
\doihref{10.1007/JHEP12(2013)062}
{JHEP \textbf{12} (2013), 062}.
%\arxivlink{1307.6546}.


\bibitem{Butter:2014xxa}
D.~Butter, S.~M.~Kuzenko, J.~Novak and G.~Tartaglino-Mazzucchelli,
\doihref{10.1007/JHEP02(2015)111}
{JHEP \textbf{02} (2015), 111}.
%\arxivlink{1410.8682}.


\bibitem{Kuzenko:2015jda}
S.~M.~Kuzenko, J.~Novak and G.~Tartaglino-Mazzucchelli,
\doihref{10.1007/JHEP09(2015)081}
{JHEP \textbf{09} (2015), 081}.
%\arxivlink{1506.09063}.


\bibitem{Butter:2017jqu}
D.~Butter, J.~Novak and G.~Tartaglino-Mazzucchelli,
\doihref{10.1007/JHEP05(2017)133}
{JHEP \textbf{05} (2017), 133}.
%\arxivlink{1701.08163}.


\bibitem{Novak:2017wqc}
J.~Novak, M.~Ozkan, Y.~Pang and G.~Tartaglino-Mazzucchelli,
\doihref{10.1103/PhysRevLett.119.111602}
{Phys. Rev. Lett. \textbf{119} (2017) no.11, 111602}.
%\arxivlink{1706.09330}.



\bibitem{Bobev:2020egg}
N.~Bobev, A.~M.~Charles, K.~Hristov and V.~Reys,
\doihref{10.1103/PhysRevLett.125.131601}
{Phys. Rev. Lett. \textbf{125} (2020) no.13, 131601}.
%\arxivlink{2006.09390}.


\bibitem{Bobev:2021oku}
N.~Bobev, A.~M.~Charles, K.~Hristov and V.~Reys,
\doihref{10.1007/JHEP08(2021)173}
{JHEP \textbf{08} (2021), 173}.
%\arxivlink{2106.04581}.


\bibitem{Bobev:2021qxx}
N.~Bobev, K.~Hristov and V.~Reys,
\doihref{10.1007/JHEP04(2022)088}
{JHEP \textbf{04} (2022), 088}.
%\arxivlink{2112.06961}.


\bibitem{Liu:2022sew}
J.~T.~Liu and R.~J.~Saskowski,
\doihref{10.1007/JHEP05(2022)171}
{JHEP \textbf{05} (2022), 171}.
%\arxivlink{2201.04690}.


\bibitem{Cassani:2022lrk}
D.~Cassani, A.~Ruip{\'e}rez and E.~Turetta,
\doihref{10.1007/JHEP11(2022)059}
{JHEP \textbf{11} (2022), 059}.
%\arxivlink{2208.01007}.


\bibitem{Gold:2023ymc}
G.~Gold, J.~Hutomo, S.~Khandelwal, M.~Ozkan, Y.~Pang and G.~Tartaglino-Mazzucchelli,
\doihref{10.1103/PhysRevLett.131.251603}
{Phys. Rev. Lett. \textbf{131} (2023) no.25, 251603}.
%\arxivlink{2309.07637}.


\bibitem{Gold:2023ykx}
G.~Gold, J.~Hutomo, S.~Khandelwal and G.~Tartaglino-Mazzucchelli,
\doihref{10.1007/JHEP07(2024)221}
{JHEP \textbf{07} (2024), 221}.
%\arxivlink{2311.00679}.


\bibitem{Ozkan:2024euj}
M.~Ozkan, Y.~Pang and E.~Sezgin,
\doihref{10.1016/j.physrep.2024.07.002}
{Phys. Rept. \textbf{1086} (2024), 1-95}.
%\arxivlink{2401.08945}.



\bibitem{Ma:2024ynp}
L.~Ma, P.~J.~Hu, Y.~Pang and H.~Lu,
%``Effectiveness of Weyl gravity in probing quantum corrections to AdS black holes,''
\doihref{10.1103/PhysRevD.110.L021901}{Phys. Rev. D \textbf{110}, no.2, L021901 (2024)}.
%[arXiv:2403.12131 [hep-th]].

\bibitem{Hristov:2025ygn}
K.~Hristov, S.~Khandelwal, Y.~Pang and G.~Tartaglino-Mazzucchelli,
\arxivlink{2511.22546}.


\bibitem{Kuzenko:2008ep}
S.~M.~Kuzenko, U.~Lindstrom, M.~Rocek and G.~Tartaglino-Mazzucchelli,
\doihref{10.1088/1126-6708/2008/09/051}
{JHEP \textbf{09} (2008), 051}.
%\arxivlink{0805.4683}.


\bibitem{Galperin:2001seg}
A.~S.~Galperin, E.~A.~Ivanov, V.~I.~Ogievetsky and E.~S.~Sokatchev,
\doihref{10.1017/CBO9780511535109}
{\emph{Harmonic superspace}, Cambridge University Press, 2007}.


\bibitem{Butter:2015nza}
D.~Butter,
\doihref{10.1007/JHEP03(2016)107}
{JHEP \textbf{03} (2016), 107}.
%\arxivlink{1508.07718}.




\bibitem{Cremmer:1984hj}
E.~Cremmer, C.~Kounnas, A.~Van Proeyen, J.~P.~Derendinger, S.~Ferrara, B.~de Wit and L.~Girardello,
\doihref{10.1016/0550-3213(85)90488-2}
{Nucl. Phys. B \textbf{250}, 385-426 (1985)}.


\bibitem{deWit:2001bk}
B.~de Wit, M.~Rocek and S.~Vandoren,
\doihref{10.1016/S0370-2693(01)00636-0}
{Phys. Lett. B \textbf{511} (2001), 302-310}.
%\arxivlink{hep-th/0104215}.


\bibitem{Kuzenko:2006nw}
S.~M.~Kuzenko,
\doihref{10.1016/j.physletb.2006.05.054}
{Phys. Lett. B \textbf{638} (2006), 288-291}.
%\arxivlink{hep-th/0602050}.


\bibitem{Butter:2014xua}
D.~Butter,
\doihref{10.1007/JHEP06(2015)161}
{JHEP \textbf{06} (2015), 161}.
%\arxivlink{1410.3604},






\end{thebibliography}
\end{document}